\documentclass[letterpaper,preprintnumbers,prd,nofootinbib,nobibnotes,showpacs]{revtex4-2}
\usepackage[colorlinks=true, linkcolor=blue, citecolor=blue, urlcolor=blue]{hyperref}
\usepackage{graphicx}
\usepackage{subfig}
\usepackage{float}
\usepackage{color}
\usepackage{amsfonts}
\usepackage{mathrsfs}
\usepackage{epsfig}
\usepackage{dcolumn}
\usepackage{amsmath}
\usepackage{color}

\begin{document}
	\renewcommand{\baselinestretch}{1.3}
	\newcommand\beq{\begin{equation}}
		\newcommand\eeq{\end{equation}}
	\newcommand\beqn{\begin{eqnarray}}
		\newcommand\eeqn{\end{eqnarray}}
	\newcommand\nn{\nonumber}
	\newcommand\fc{\frac}
	\newcommand\lt{\left}
	\newcommand\rt{\right}
	\newcommand\pt{\partial}

\allowdisplaybreaks

\title{Generalized accelerating Kerr-Newman-NUT-de Sitter black holes }
    \author{Changjun Gao}\email{gaocj@nao.cas.cn}
\affiliation{Key Laboratory of Computational Astrophysics, National Astronomical Observatories, Chinese
	Academy of Sciences, Beijing 100012, China}
\affiliation{University of Chinese Academy of Sciences, Beijing 100049, China}

\begin{abstract}
We obtain rotating black hole solutions to the Einstein-Maxwell equations with the cosmological constant. There are eight parameters in the solutions. They are the physical mass $M$, the electric charge $Q$, the specific angular momentum $a$, the acceleration $\alpha$, the cosmological constant $\lambda$ and other three parameters, $S_1\;, S_2$ and $S_3$. We show that $S_3$ and $S_1$ are actually the well-known NUT charge and the rescaled parameter $C$, respectively, for the Kerr-NUT spacetime. As for $S_2$, it is a new parameter although it has the same physical meaning of $C$ in the weak field limit. In the case of static limit, $S_2$ describes a conical defect along the axis of revolution, which corresponds to a cosmic string of tension. Therefore, the solutions extend the accelerating Kerr-Newman-NUT-de Sitter spacetime. 
\end{abstract}

%11.10.Kk Field theories in dimensions other than four (see also 04.50.-h Higher-dimensional gravity and other theories of gravity; 04.60.Kz Lower dimensional models; minisuperspace models in general relativity and gravitation)

%04.50.Kd 	Modified theories of gravity

\maketitle


%%%%%%%%%%%%%%%%%%%%%%%%%%%%%%
\section{Introduction}
It is of great importance to find exact solutions to the Einstein equations in order to study the geometry of spacetime and the flow of energy and momentum. These solutions help us understand gravity and the evolution of the Universe. They also provide us many fascinating and unexpected physical objects, such as black holes, wormholes, gravitational waves and cosmic expansion. Here we do not think it is necessary to give a detailed review on exact solutions to Einstein's equations because an elegant presentation has been made in the classic books \cite{griffiths2009exact,stephani2009exact}. Where we start from is only the Kerr-NUT black hole solutions. 

The vacuum Einstein equations gives the metric of Kerr-NUT black holes  
\begin{eqnarray}\label{eq:kn}
ds^2&=&-\frac{\Delta_1}{\Sigma_1}[dt+(2n\cos\theta +2Cn-a\sin^2\theta)d\phi]^2+\frac{\Sigma_1}{\Delta_1}dr^2\nonumber\\&&
 +\frac{\sin^2\theta}{\Sigma_1}[adt-(r^2+a^2+n^2-2anC)d\phi]^2+\Sigma_1d\theta^2\;, 
 \end{eqnarray}
  with
\begin{eqnarray} 
 \Delta_1 =r^2 +a^2-2Mr-n^2\;,\ \ \ \ 
 \Sigma_1=r^2 +\left(n+a\cos\theta\right)^2\;,
\end{eqnarray}
where $M$ is the physical mass, $n$ the NUT charge, $a$ the specific angular momentum and parameter $C$ specifies different configurations for the strings \cite{misner1963flatter,demianski1966combined,manko2006singular}. 
The presence of  NUT charge causes string-like defects \cite{misner1963flatter}. These defects are named after Misner strings. The parameter $C$ was firstly introduced by
 Manko and Ruiz \cite{manko2006singular}. Observing the metric, we find that $C$ can be rescaled to $\tilde{C}\equiv{Cn}$. In other words, $n$ can be absorbed by $C$ such that $C$ can be assigned with discrete values, $C=0,\pm{1}$. Different discrete values of $C$ leads to different configurations for the Misner strings. In specific, $C=1$ gives the string at the north 
pole while the south pole is regular. On the contrary, $C=-1$ gives the string at the south  pole while the north pole is regular. When $C=0$,  two strings are symmetrically located at the north and south poles.

The Kerr-NUT spacetime is the solution to the vacuum Einstein equations. One of the generalizations of this spacetime is the Kerr-Newman-NUT–de Sitter space-time. It satisfies the Einstein–Maxwell equations with the cosmological constant. Thus two additional parameters, electric (or magnetic) charge and the cosmological constant are included in the solution. This class of generalizations was completed by Pleba$\acute{\mathrm{n}}$ski \cite{plebanski1975class}. An even larger class in which an acceleration parameter is introduced, was found by  Pleba$\acute{\mathrm{n}}$ski and Demia$\acute{\mathrm{n}}$ski \cite{plebanski1976rotating}.  Detailed  discussions on the Pleba$\acute{\mathrm{n}}$ski-Demia$\acute{\mathrm{n}}$ski metrics can be  found in the book by Griffiths and Podolsk$\acute{\mathrm{y}}$\cite{griffiths2009exact} (see also  Stephani et al. \cite{stephani2009exact}).

In this paper, we report a new solution to the Einstein-Maxwell equations with the cosmological constant. We show that it is a generalization of accelerating Kerr-Newman-NUT-de Sitter spacetime with a new parameter $S_2$. In the weak field limit, the parameter $S_2$ has the same physical meaning of $C$. In the case of static spacetime, $S_2$ describes the tension of cosmic strings.

%%%%%%%%%%%%%%%%%%%%%%%%%%%%%%
\section{solutions}

We find the following line element

\begin{eqnarray}\label{eq:sol}
ds^2&=&\frac{1}{\Omega^2}\left\{-\frac{\Sigma-\delta{a^2}\sin^2\theta}{\Delta}dt^2
+\frac{\Delta}{\Sigma}dr^2
+\frac{\Delta}{\delta}d\theta^2-\frac{2a\sin^2\theta\left[\delta\left(r^2+a^2\right)-\Sigma\right]}{\Delta}dtd\phi\right.\nonumber\\&&\left.+\frac{\left[\delta\left(r^2+a^2\right)^2-\Sigma{a}^2\sin^2\theta\right]\sin^2\theta}{\Delta}{d}\phi^2\right\}\;,
\end{eqnarray}
with
\begin{eqnarray}
\Delta&=&r^2+a^2\cos^2\theta\;,\ \ \ \ \ \ \ \Omega=1-\alpha{r}\cos\theta\;,\nonumber\\
\delta&=&1+\frac{1}{3}\lambda{a^2}\cos^2{\theta}+\frac{S_1+S_3\cos\theta-S_2\cos^2\theta}{\sin^2\theta}+\frac{2\alpha{M}\cos^3\theta}{\sin^2\theta}-\frac{\alpha^2(a^2S_1+a^2+Q^2)\cos^4\theta}{\sin^2\theta}\;,\nonumber\\  
\Sigma&=&\left(r^2+a^2\right)\left(1-\frac{1}{3}\lambda{r}^2\right)-2Mr+Q^2+S_1a^2+S_2r^2-S_3\alpha{r}^3-\alpha^2\left(1+S_1\right)r^4\;,\ \ \ \ \ \  
\end{eqnarray}
and the Maxwell field
\begin{eqnarray}
A_{0}=\frac{Qr}{\Delta}\;,\ \ \ \ \ \ \  A_{3}=-a\sin^2\theta{A_0}\;,  
\end{eqnarray}
satisfy the Einstein-Maxwell equations with the cosmological constant
\begin{equation}
	G_{\mu\nu}=-2F_{\mu\alpha}F_{\nu}^{\ \alpha}+\frac{1}{2}g_{\mu\nu}F_{\alpha\beta}F^{\alpha\beta}+\lambda{g_{\mu\nu}}\;, \ \ \ \ \ \ \    \nabla_{\mu}F^{\mu\nu}=0\;,
\end{equation}
where $M,a,\alpha, Q,\lambda,S_1,S_2,S_3$ are constants. It is apparent $S_i$ are dimensionless constants. When $S_1=S_2=S_3=0$, the metric reduces to the accelerating  Kerr-Newman-de Sitter spacetime. Therefore, $M,Q,a,\alpha$ and $\lambda$ represent the physical mass, electric charge, specific angular momentum, acceleration  and the cosmological constant, respectively. 
It is noticeable that the Maxwell field here is the same as that in Kerr-Newman-de Sitter spacetime.  Same as the Kerr-Newman-de Sitter spacetime, there is a singular circular ring in the center of the black holes, i.e. $r=0$ and $\theta=\frac{\pi}{2}$. 

Then what are the physical meanings of $S_i$? In order to figure out this point, we make coordinate transformations. For our purposes, we are concerned with the case of
\begin{eqnarray}
	\lambda=0\;,\ \ \ \ Q=0\;,
\end{eqnarray}
in the line element Eq.~(\ref{eq:sol}).
We parametrize $S_i$ as follows
\begin{eqnarray}
	S_1=-\frac{n^2-B^2+a^2b^2}{a^2b^2}\;,\ \ \ \ S_2=-1+b^2\;, \ \ \ \ S_3=-\frac{2n}{a}\;,
\end{eqnarray}
where $n,b,B$ are constants. Then the parameters $S_i$ are replaced with $n,b$ and $B$.
Now we make coordinate transformation  $\theta\longrightarrow{\tilde{\theta}}$ via
\begin{eqnarray}
	\tilde{\theta}=\int{\frac{1}{\sqrt{\delta}}d\theta}\;,
\end{eqnarray}

namely, 

\begin{eqnarray}
	\theta=\arccos\frac{-n+B\sin^2(b\tilde{\theta})}{ab^2}\;.
\end{eqnarray}

Making coordinate transformation once more, $\tilde{\theta}\longrightarrow{\tilde{\tilde{\theta}}}$ 

\begin{eqnarray}
	\tilde{\theta}=\frac{\tilde{\tilde{\theta}}}{b}+\frac{\pi}{2b}\;,
\end{eqnarray}

and performing the following variable substitutions

\begin{eqnarray}
	\tilde{\tilde{\theta}}\longrightarrow{\theta}\;,\ \ \ \ a^2\longrightarrow{-2Can+n^2+a^2}\;,\ \ \ \  B\longrightarrow{-a}\;,
\end{eqnarray}

we eventually find that the metric functions become

\begin{eqnarray}\label{eq:cnb}
&&g_{00}=-\frac{(-n^2+a^2\cos^2\theta-2Mrb^2+r^2b^4)b^2}{r^2b^4+(n+a\cos\theta)^2}\;,\nonumber\\
	&&g_{11}=\frac{r^2b^4+(n+a\cos\theta)^2}{b^2(-2Mrb^2-n^2+a^2+r^2b^4}\;,\nonumber\\
	&&g_{22}=\frac{r^2b^4+(n+a\cos\theta)^2}{b^6}\;,\nonumber\\
	&&g_{03}=\frac{a^2\cos^2\theta(n^2+b^4a^2+b^4n^2+2Mrb^2-2b^4anC-a^2)}{b^2[r^2b^4+(n+a\cos\theta)^2]}
	\nonumber\\&&-\frac{2an(a^2-n^2-2Mrb^2+r^2b^4)\cos\theta}{b^2[r^2b^4+(n+a\cos\theta)^2]}\nonumber\\&&+\frac{(2an^3C-n^4-a^2n^2-r^2a^2-n^2r^2)b^4+2Mrb^2n^2-a^2n^2+n^4}
	{b^2[r^2b^4+(n+a\cos\theta)^2]}\nonumber\\&&+\frac{rb^4[(rn^2+ra^2-2ranC)b^2-2Ma^2+4ManC-2Mn^2]}{r^2b^4+(n+a\cos\theta)^2}
	\;,\nonumber\\
	&&g_{33}=-\frac{a^3\cos^3\theta(a^2-n^2-2Mrb^2+r^2b^4)(a\cos\theta+4n)}{b^6[r^2b^4+(n+a\cos\theta)^2]}
	\nonumber\\&&-\frac{a^2\cos^2\theta[(4a^3nC-4an^3C-2a^4+6n^2r^2+2n^4)b^4-12Mrb^2n^2-6n^4+6a^2n^2]}{b^6[r^2b^4+(n+a\cos\theta)^2]}
	\nonumber\\&&-\frac{a^2\cos^2\theta\left\{[(a^2+n^2)^2+r^4+4anC(anC-a^2-n^2)]b^2+4Mr(n^2+a^2-2anC)\right\}}{r^2b^4+(n+a\cos\theta)^2}\nonumber\\&&
	-\frac{a\cos\theta[(4n^3r^2+8a^3n^2C-8n^4aC+4n^5-4a^4n)b^4-8Mrb^2n^3+4a^2n^3-4n^5]}{b^6[r^2b^4+(n+a\cos\theta)^2]}\nonumber\\&&
	+\frac{4nar\cos\theta((rn^2+ra^2-2ranC)b^2-2Ma^2+4ManC-2Mn^2)}{r^2b^4+(n+a\cos\theta)^2}\nonumber\\&&
	-\frac{(r^2n^4-2n^2a^4+2n^6+4a^3n^3C-4n^5aC)b^4-2Mrb^2n^4+n^4a^2-n^6}{b^6[r^2b^4+(n+a\cos\theta)^2]}\nonumber\\&&
	-\frac{-8Mrn^3aC+4a^2Mrn^2+4Mrn^4}{r^2b^4+(n+a\cos\theta)^2}\nonumber\\&&
	-\frac{b^2[(-2r^2-n^2)a^4+(4r^2Cn+4n^3C)a^3+(-r^4-2n^4-4n^4C^2-4n^2r^2)a^2]}{r^2b^4+(n+a\cos\theta)^2}\nonumber\\&&
	-\frac{b^2[(4r^2n^3C+4n^5C)a-2r^2n^4-n^6]}{r^2b^4+(n+a\cos\theta)^2}\nonumber\\&&
	-\frac{rb^4(-2a^4M-8Ma^2n^2C^2+8Man^3C+8Ma^3nC-4Ma^2n^2-2Mn^4)}{r^2b^4+(n+a\cos\theta)^2}\nonumber\\&&
	-\frac{r^2b^6(a^4+4a^2n^2C^2-4an^3C-4a^3nC+2a^2n^2+n^4)}{r^2b^4+(n+a\cos\theta)^2}\;.
\end{eqnarray}

We see the metric functions are rather complicated in this coordinate system. On the contrary, the original expressions in Eq.~(\ref{eq:sol}) are relatively simple. But Eq.~(\ref{eq:cnb}) helps us understand the physical meaning of $S_i$. Of course, Eq.~(\ref{eq:cnb}) now solves the vacuum Einstein equations because both the electric charge $Q$ and cosmological constant $\lambda$ vanish now. We have checked that Eq.~(\ref{eq:cnb}) does satisfy the vacuum Einstein equations.  Compared with the Kerr-NUT solutions Eq.~(\ref{eq:kn}), one extra parameter $b$ emerges in Eq.~(\ref{eq:cnb}) although  they also satisfy the vacuum Einstein equations. Therefore, it is an extension of Kerr-NUT solution.  For specific, when $b=1$, the solution reduces to exactly the Kerr-NUT spacetime. $n$ is actually the NUT charge and the parameter $C$ specifies different configurations for the strings. Now we realize that $S_1$ is actually the rescaled parameter $C$ and $S_3$ labels the NUT charge. As for $b$ or $S_2$, it is a new parameter. To gain some understanding of  the physical meaning of $b$, we let 
$n=0$ and $C=0$ in Eq.~(\ref{eq:cnb}). Then the solutions can be simplified to 
\begin{eqnarray}
	g_{00}&=&-\frac{a^2\cos^2\theta-2Mr+r^2}{a^2\cos^2\theta+r^2}\;,\nonumber\\
	g_{11}&=&\frac{a^2\cos^2\theta+r^2}{-2Mr+a^2+r^2}\;,\nonumber\\
	g_{22}&=&a^2\cos^2\theta+r^2\;,\nonumber\\
	g_{03}&=&\frac{a[(-a^2+a^2b^4+2Mr)\cos^2\theta+b^4r^2-2Mrb^4-r^2)]}{a^2\cos^2\theta+r^2}\;,\nonumber\\
	g_{33}&=&-\frac{a^2\cos^4\theta(-2Mr+a^2+r^2)}{r^2+a^2\cos^2\theta}
	\nonumber\\&&-\frac{(a^4b^8+r^4-2a^4b^4+4Mra^2b^4)\cos^2\theta}{r^2+a^2\cos^2\theta}
	\nonumber\\&&-\frac{a^2b^8r^2-2Mra^2b^8-2b^4a^2r^2-r^4}{r^2+a^2\cos^2\theta}\;.
\end{eqnarray}
It is now the extension of Kerr spacetime with a new parameter $b$ and solves the vacuum Einstein equations. When $b=1$, they go back to the case for Kerr spacetime. We notice that due to the presence of $b$, the spacetime is no longer asymptotically flat in space. In order to grasp some physical significance of $b$, we make series expansion of the metric in the weak field limit. Then we obtain 
  \begin{eqnarray}
ds^2&=&-\left(1-\frac{2M}{r}\right)dt^2+\left(1-\frac{2M}{r}\right)^{-1}dr^2+r^2d\Omega_2^2\nonumber\\&&
+2\left[\left(ab^4-a\right)-\frac{2M(ab^4-a\cos^2\theta)}{r}\right]dtd\phi\;.
\end{eqnarray}
 As expected, when $b=1$, it is exactly the Lense-Thirring metric which is the weak field limit of Kerr spacetime. On the other hand, if we make series expansion of Kerr-NUT solution with $n=0$ and $\tilde{C}\equiv{nC}\neq{0}$, we would obtain 
 \begin{eqnarray}
	ds^2&=&-\left(1-\frac{2M}{r}\right)dt^2+\left(1-\frac{2M}{r}\right)^{-1}dr^2+r^2d\Omega_2^2\nonumber\\&&
	+2\left[-2\tilde{C}+\frac{2M(2\tilde{C}-a\sin^2\theta)}{r}\right]dtd\phi\;.
\end{eqnarray}
Comparing the two metrics, we conclude that if we let 
 \begin{eqnarray}
b^4=1-\frac{2\tilde{C}}{a}\;,
\end{eqnarray}
$b$ would have the same physical meaning of rescaled factor $C$, in  the limit of weak field approximation. Therefore, the physical meaning of $b$ is to make spacetime no longer asymptotically flat.

%%%%%%%%%%%%%%%%%%%%%%%%%%%%%%
\section{Understanding $S_2$ in the case of static spacetime}

In this section, let's understand the effect of $S_2$ when $M\neq{0}$ and $S_2\neq{0}$. We find the metric Eq.~(\ref{eq:sol}) becomes after coordinate transformations
\begin{eqnarray}\label{eq:conical}
	ds^2=-\left[1-\frac{2M}{\left(1+S_2\right)r}\right]dt^2+\left[1-\frac{2M}{\left(1+S_2\right)r}\right]^{-1}dr^2+r^2\left[d\chi^2+\left(1+S_2\right)\sin^2\chi{d}\phi^2\right]\;.
\end{eqnarray}
We see when $S_2\neq{0}$, both the symmetry and the physical mass of spacetime are changed. Therefore, $S_2$ is a physical quantity used to measure the degree to which a black hole deviates from sphericity and the degree to which the physical mass $m$ deviates from the Comar mass $m=\frac{M}{1+S_2}$. In fact, the metric describes a Schwarzschild black hole with a cosmic string passing through it. This spacetime is firstly studied by Aryal, Ford and Vilenkin \cite{aryal1986cosmic}.  If
$S_2\neq{0}$, there would be a conical defect along the axis of revolution, which corresponds to a cosmic string of tension
 \begin{eqnarray}
	\mu=\frac{\delta_0}{8\pi}=\frac{1}{4}\left(1-{\sqrt{1+S_2}}\right)\;,
\end{eqnarray}
where $\delta_0$ is the conical deficit.  This interpretation of tension is justified by studying the
 equations of motion of a cosmic string vortex in the background of a black hole spacetime
 where Eq.~(\ref{eq:conical}) was obtained as the background spacetime outside the string core \cite{achucarro1995abelian}. 

In particular, when
\begin{eqnarray}
	M\neq{0}\;,\ \ \ \ \ \ \  S_2=-1\;,
\end{eqnarray}
we obtain after coordinate transformation
\begin{eqnarray}
	ds^2=\frac{2M}{r}dt^2-\frac{r}{2M}dr^2+r^2\left(dx^2+dy^2\right)\;.
\end{eqnarray}
It is a planar solution which is studied by Chen and Liu \cite{chen2025taub} recently.  If we perform coordinate transformation once more, $r=\frac{1}{2}\left(36M\tau^2\right)^{\frac{1}{3}}$ and $t=z$, the metric would become
 \begin{eqnarray}
 	ds^2=-d\tau^2+\frac{2}{3}\left(\frac{6M^2}{\tau^2}\right)^{\frac{1}{3}}dz^2+\frac{3}{2}\left(6M^2\tau^4\right)^{\frac{1}{3}}\left(dx^2+dy^2\right)\;.
 \end{eqnarray}
 It is the Kasner plane solution \cite{kasner1921geometrical}. The solution shows the universe expands in the $x-y$ plane and contracts in the $z$ direction. With the increasing of cosmic time, the universe is compressed in the $x-y$ plane eventually. To summarize, in the case of static spacetime, $S_2$ describes the tension of cosmic strings.

%%%%%%%%%%%%%%%%%%%%%%%%%%%%%%
\section{thermodynamics}

In this section, we turn to the horizon thermodynamics of black holes. The spacetime of  Eq.~(\ref{eq:sol}) is stationary and axisymmetric, corresponding 
to the Killing vector
\begin{equation}
K^{\mu}\equiv\xi^{\mu}+\Omega_{H}\varphi^{\mu}\;,
\end{equation}
where $\xi^{\mu}\equiv{\partial_{t}}$ and $\varphi^{\mu}\equiv{\partial_{\phi}}$ and $\Omega_{H}$ is the angular velocity of the ``Zero Angular Momentum 
Observer'' (ZAMO) evaluated on the horizons
\begin{equation}
	\Omega_{H}\equiv\omega\Bigg|_{r=r_i}=-\frac{g_{03}}{g_{33}}\Bigg|_{r=r_i}\;.
\end{equation}
The Killing horizons admitted by Killing vector are governed by 
\begin{equation}\label{eq:hor1}
\Sigma=0\;. 
\end{equation}

This is a quartic algebraic equation. In general, it has four roots. We factorize Eq.~(\ref{eq:hor1}) with four positive roots $c_1,c_2,c_3,c_4$ and then we obtain 

\begin{eqnarray}\label{roots}
M&=&M(c_1,c_2,c_3,c_4,\alpha, S_3)\;,\ \ \ \ \ \ \ \ Q=Q(c_1,c_2,c_3,c_4,\alpha,S_2,S_3)\;,\nonumber\\
S_1&=&S_1(c_1,c_2,c_3,c_4,\alpha, S_2, S_3)\;,\ \ \ \ \lambda=\lambda(c_1,c_2,c_3,c_4,\alpha,S_2,S_3)\;.
\end{eqnarray}
As an example, if we want four roots, 
\begin{eqnarray}
c_1=1\;, \ \ c_2=2\;, \ \ c_3=3\;, \ \ c_4=4\;, 
\end{eqnarray}
we  would obtain 
\begin{eqnarray}
M=1\;, \ \ \alpha=1\;, \ \ Q=\frac{\sqrt{7}}{10}\;, \ \ \lambda=-\frac{54}{5}\;, \ \ S_1=\frac{64}{25}\;,\ \ S_2=-\frac{1}{2}\;,\ \ S_3=\frac{2}{5}\;,\ \ a=\frac{1}{2}\;.
\end{eqnarray}
$c_1$ and $c_2$ are for black hole horizons while $c_3$ and $c_4$ the cosmic horizons.

The angular velocity has the following limits

\begin{eqnarray}
	&&\lim_{r\rightarrow{r_i}}\omega\equiv\Omega_{H}=\frac{a}{a^2+r^2}\Bigg|_{r=r_i}\;,\nonumber\\  &&\lim_{r\rightarrow{+\infty}}\omega\equiv{\Omega_{\infty}}\nonumber\\&&={(\lambda+3\alpha^2+3\alpha^2S_1)a\sin^2\theta}\left[3S_1+3a^2\alpha^2S_1+3a^2\alpha^2+a^2\lambda+3+3S_3\cos\theta
    \right.\nonumber\\&&\left.-3Q^2\alpha^2\cos^4\theta+6\alpha{M}\cos^3\theta-(3+6a^2\alpha^2+3S_2+a^2\lambda+6a^2\alpha^2S_1)
    \cos^2\theta\right]^{-1}\;.
\end{eqnarray}
The fact that angular velocity in infinity $\Omega_{\infty}$ is dependent on the angle $\theta$ is a salient feature of these rotating black
holes, different from the case of Kerr-de Sitter black hole, where $\Omega_{\infty}=const$ \cite{sekiwa2006thermodynamics}. It is interesting that $\Omega_{\infty}$ vanishes on the north and south poles. The surface gravity $\kappa$ is defined as the ratio of $\xi^{\mu}\nabla_{\mu}\xi^{{\nu}}$ to $\xi^{\nu}$ and the temperature of horizons, $T\equiv{\kappa}/{2\pi}$  are
\begin{equation}\label{eq:tem}
T=\frac{(9\alpha^2S_1+9\alpha^2+3\lambda)r^4+6\alpha{S}_3{r}^3+(a^2\lambda-3S_2-3)r^2+3S_1a^2+3a^2+3Q^2}{12\pi{r}(r^2+a^2)}\Bigg|_{r=r_i}\;, 
\end{equation}
while the Bekenstein-Hawking entropy is 
\begin{equation}
S=\frac{A}{4}=\frac{1}{4}\int_{0}^{\pi}{d\theta}\int_{0}^{2\pi}d\phi\sqrt{g_{22}g_{33}}\Bigg|_{r=r_i}=\frac{\pi\left(r^2+a^2\right)}{1-\alpha^2{r}^2}\Bigg|_{r=r_i}\;.
\end{equation}
We notice that the temperature here is named after Gibbsian temperature by Cvetic et al \cite{cvetivc2018killing}. They show that if one regards the Christodoulou and Ruffini
formula for the total energy or enthalpy as defining the Gibbs surface, then the rules of Gibbsian thermodynamics imply that negative temperatures arise inevitably on some horizons, as does the conventional form of the first law. We will see this point sooner. 

We observe that if one of the horizons locates at 
\begin{equation}\label{eq:ralpha}
r_c=\frac{1}{\alpha}\;,
\end{equation}
its Bekenstein-Hawking entropy would be infinite. Why is it infinite? The reason is that the spacetime is no longer asymptotically flat which is indicated in the previous section. So the area of the horizon is no longer finite although its radius is. Can the spacetime have a horizon with the radius $r_c$? The answer is yes. Substituting  Eq.~(\ref{eq:ralpha}) into the equation of horizons, Eq.~(\ref{eq:hor1}), we can solve for the mass $M$. What we want is merely $M>0$. We find it is indeed the case.  As for the temperatures, we find 

\begin{equation}
T_1<0\;,\ \  T_2>0\;,\ \ T_3<0\;,\ \ T_4>0\;,
\end{equation}
where $T_1,T_2,T_3,T_4$ represent the temperatures of the horizons from the inside to the outside.

We can arrange the the horizons to coincide as follows  
\begin{equation}
r_1=r_2<r_3=r_4\;.
\end{equation}
Then it leeds to 
\begin{equation}
T_1=T_2=0\;,\ \  T_3=T_4=0\;.
\end{equation}
Namely, we obtain a spacetime with two horizons, both with vanishing temperature.  Carballo-Rubio et al \cite{carballo2022regular} showed that,  if one  makes attempt to let the surface gravity of  inner horizons vanishes, then the exponential growth character of mass inflation instability is not present. Therefore, the above two-horizon spacetime does not suffer from the mass-inflation problem.

The electric potential $V_{0}$ on the horizons are  
\begin{equation}
	V_{0}=\frac{Qr}{r^2+a^2}\Bigg|_{r=r_i}\;.
\end{equation}
The thermodynamics of Kerr-NUT black holes are discussed in \cite{bordo2019first,frodden2022first}.  A consistent thermodynamics for the Kerr–Newman-NUT black hole is got by Yang et al. \cite{yang2023first}. Other consistent thermodynamics for black hole with NUT charge have been formulated in \cite{hennigar2019thermodynamics,bordo2019thermodynamics,durka2022first,clement2020smarr,wu2019thermodynamical,chen2019general,abbasvandi2021thermodynamics,liu2022thermodynamics,gregory2017accelerating}.
We have tried our best to construct horizon thermodynamics for the generalized Kerr-Newman-NUT-de Sitter black holes. However, we have not been able to achieve this goal. Only in the following situations, 
\begin{equation}
\alpha=0\;,\ \ \ \ \lambda=0\;,\ \ \ \ S_2=S_1\;,
\end{equation}
we find the Smarr formula %
\begin{equation}\label{eq:smar}
	M=2TS+2\Omega_{H}J+QV_{0}\;,
\end{equation}
and the first law of thermodynamics 
\begin{eqnarray}\label{eq:fir}
	dM=TdS+\Omega_{H}dJ+V_{0}dQ\;,
\end{eqnarray}
are satisfied. Here $J=Ma$ is the angular momentum of the black holes. We think the main reason for future success might lie in dealing with the angular momentum at infinity $\Omega_{\infty}$ properly because this term is expected to enter into the Smarr formula and the first law of thermodynamics following the studies of thermodynamics on Kerr-Newman-anti-de Sitter black holes \cite{caldarelli2000thermodynamics}. We know $\Omega_{\infty}=const$ in the Kerr-Newman-anti-de Sitter spacetime. However, in the generalized Kerr-Newman-NUT-de Sitter (or anti-de Sitter) spacetime, $\Omega_{H}$ is angle-dependent and it is not a constant.

%%%%%%%%%%%%%%%%%%%%%%%%%%%%%%
\section{Conclusions}

In conclusion, we obtain exact solutions to the Einstein-Maxwell equations with the cosmological constant. The solutions have eight parameters and generalizes the accelerating Kerr-Newman-NUT-de Sitter spacetime with a new parameter $S_2$. Compared with familiar expressions for Kerr-Newman-NUT-de Sitter spacetime, our metric functions are considerable simple.  In the limit of weak field approximation, the new parameter $S_2$ has the same physical meaning as $C$ for the Kerr-NUT spacetime. On the other hand, in the case of static spacetime, $S_2$ describes the tension of a cosmic string. We find that the spacetime can have four horizons in general, two black hole horizons and two cosmic horizons. To our knowledge, this kind of four-horizon structure never occur in the accelerating Kerr-Newman-NUT-de Sitter spacetime.      

We make an exploration of the thermodynamics of the generalized Kerr-Newman-NUT-de Sitter black holes. What we are able to do is constructing the Smarr formula and the first law of thermodynamics only in the case of $\alpha=0, S_2=S_1$ and $\lambda=0$. Since the term of angular momentum at infinity are expected to play a role in the thermodynamics formulas, one of the huge obstacles towards to the angle-dependent $\Omega_{\infty}$. It is also amusing that $S_3$ does not appear in our thermodynamics formulas. As shown before, $S_3$ corresponds to the rescaled NUT charge and the NUT charge always plays important role in the first law of thermodynamics (for instance, see \cite{yang2023first}). We think the reason for this comes from the difference of coordinate systems. Finally, an interesting observation is that the spacetime can have two degenerate horizons, both with vanishing temperatures. Following the studies made by Carballo-Rubio et al \cite{carballo2022regular}, the mass-inflation phenomenon does not happen in this kind of two-horizon spacetime.

\section*{ACKNOWLEDGMENTS}

The work is partially supported by the Special Exchange Program of CAS, National Key RD Program of China grants No.
2022YFF0503404, No. 2022SKA0110100 and the Central Guidance for Local Science and Technology Development Fund Project with Grand No. 2024ZY0113.

\bibliography{reference.bib}

\end{document}